\documentstyle[a4,11pt]{article}
\newcommand{\beqn}{\begin{eqnarray}}
\newcommand{\eeqn}{\end{eqnarray}}

\def\sgn{\mathop{\mathrm{sgn}}}
\def\Tr{\mathop{\mathrm{Tr}}}

\newmathalphabet{\bmabn}
\addtoversion{normal}{\bmabn}{cmr}{bx}{n}
\newmathalphabet{\bmabit}
\addtoversion{normal}{\bmabit}{cmm}{b}{it}

\begin{document}

\title{
The Third Virial Coefficient of Free Anyons
}

\author{
Jan Myrheim\thanks{
Supported by the Norwegian Research
Council for Science and the Humanities,
NAVF.}
\ and K\aa re Olaussen
\\
Institutt for fysikk, NTH,\\
N--7034 Trondheim, Norway.}

\date{October 29, 1992}

\begin{titlepage}

\vspace*{\fill}

\centerline{\huge
The Third Virial Coefficient of Free Anyons
}

\vspace{2cm}

\centerline{\large
Jan Myrheim\footnote{
Supported by the Norwegian Research
Council for Science and the Humanities,
NAVF.}
and K\aa re Olaussen}
\centerline{\large
Institutt for fysikk, NTH,}
\centerline{\large
N--7034 Trondheim, Norway.}
\vspace{0.5cm}
\centerline{\today}
\vspace{0.5cm}
\centerline{cond-mat/9210028}
\vspace{2.5cm}

\begin{abstract}
We use a path integral representation
for the partition function of non-interacting
anyons confined in a harmonic oscillator
potential in order to prove that the third
virial coefficient of free anyons is finite,
and to calculate it numerically.
Our results together with previously
known results are consistent with
a rapidly converging Fourier series
in the statistics angle.
\end{abstract}

\vspace*{\fill}

\end{titlepage}

\section{Introduction}

The two-dimensional ``fractional statistics''
of anyons
\cite{JML:JM:idpart1,GMS:idpart1,FW:idpart1}
is not yet a particle statistics in the same
sense as Bose---Einstein or Fermi---Dirac
statistics. Not much is known even about
the gas of free anyons, beyond the second
virial coefficient $A_2(\theta)$
\cite{ASWZ:secvir,JSD:secvir}.
As usual, $\theta$ denotes the angle
defining the particle statistics.
One known exact result for the third virial
coefficient $A_3(\theta)$ is that it is symmetric
under the substitution $\theta\to\pi-\theta$.
This follows from the existence of
a certain ``supersymmetry''
transformation \cite{SenI,SenII}.
Furthermore, according to perturbation theory
there is no variation of $A_3(\theta)$
to first order in $\theta$ or $|\theta|$ near
the boson point $\theta=0$,
or in $\theta-\pi$ or $|\theta-\pi|$
near the fermion point $\theta=\pi$
\cite{SenIII,JMcC:SO:1,AC:JMcC:SO:2},
but there is a second order variation
at both points
\cite{ADV:SO:esso,MS:JJMV:IZ:a3}.

It is not entirely obvious to us
that $A_3$ should be a differentiable
function of $\theta$ at $\theta=0$ or
$\theta=\pi$, or indeed anywhere.
But if it is analytic, then it ought
to be well approximated by only a few
terms from the Fourier series
\beqn
\label{eq:A3conjecture}
A_3(\theta)={1\over 36}+
{1\over 12\pi^2}\sin^2\!\theta+
c_4\sin^4\!\theta+c_6\sin^6\!\theta
+\ldots\;.
\eeqn
This series is consistent
with periodicity and ``supersymmetry'',
and with the second order bosonic and
fermionic perturbation expansions.
We find it rather remarkable that
already the first two terms give
a good approximation to our numerical
data, with no free parameter.
If this lowest order approximation is
indeed exact, which is not totally
excluded, then there should be
hope of calculating it exactly.

Following Laidlaw and DeWitt
\cite{LDW:idpart1}, we consider
the path integral representation for
the partition function of a system of
identical particles in an external
potential. This approach
emphasizes the connection between
topology and particle statistics
\cite{LDW:idpart1,JML:JM:idpart1,APB:topstat}.
Our path integral formula is reminiscent of
(and was inspired by) the cycle
expansion for strange attractors
\cite{RA:EA:PC:cyclexp}.
The partition function of
three anyons in a harmonic oscillator
potential in the limit of zero frequency
gives the third virial coefficient
of free anyons
\cite{JMcC:SO:1,AC:JMcC:SO:2,ADV:SO:esso,MS:JJMV:IZ:a3,KO:horeg}.
The two-anyon harmonic oscillator problem is
trivial, it was solved in
Ref.\ \cite{JML:JM:idpart1}, and the solution
yields the second virial coefficient
\cite{JMcC:SO:1,AC:JMcC:SO:2}.
In the three-anyon problem the complete
energy spectrum is not known, but the lowest
levels have been calculated numerically
\cite{MS:JJMV:IZ,MVNM:JL:MB:RKB},
or perturbatively near
the boson and fermion points, and
an infinite number of exact solutions are
known with energies linearly dependent on
$\theta$.
See e.g.\ \cite{SenII,APP:xxx,JAR:FR:systNanysp}
and references given there.
The path integral representation gives
the partition function directly without
explicit knowledge of the energy levels.

One non-trivial result we obtain beyond
the numerical values, is a proof
that the third virial coefficient of free
anyons is always finite. This has so far
been known only to second order in
perturbation theory
\cite{ADV:SO:esso,MS:JJMV:IZ:a3}.
A more detailed account of our work
will be given elsewhere
\cite{JM:KO:II}.

\section{The Partition Function}

By reformulating the trace as a path
integral, we get the following formula
for the partition function of $N$ anyons
in the external potential $U$,
\beqn
\label{eq:ZNanyonI}
Z_N(\beta,\theta)=
\Tr e^{-\beta\widehat{H}}=
\sum_{\cal P}
{1\over\prod_L\,
(\nu_L!\,L^{\nu_L})}
\mathop{\int}_{{\cal C}({\cal P})}
{\cal D}({\bmabit R}(\tau))\,
e^{-{S\over\hbar}}\,
e^{-i\theta Q}.
\eeqn
Here $\widehat{H}$ is
the Hamiltonian operator.
${\bmabit R}=({\bmabit r}_1,{\bmabit r}_2,
\ldots,{\bmabit r}_N)\in{\bmabn R}^{2N}$
denotes the $N$-particle configuration,
with ${\bmabit r}_j\in{\bmabn R}^2$, and
${\bmabit R}(\tau)$ denotes
an $N$-particle path as a function
of the imaginary time $t=-i\tau$, with
$0\leq\tau\leq\hbar\beta$.
${\cal D}({\bmabit R}(\tau))$ is
the path integral measure, and
\beqn
S=\sum_{j=1}^N
\mathop{\int}_0^{\hbar\beta}d\tau\left\{
{m\over 2}\left|{d{\bmabit r}_j\over d\tau}
\right|^2+U({\bmabit r}_j)\right\}
\eeqn
is the action in imaginary time, with
$m$ the particle mass.

The trace is obtained by integration
over closed paths. With identical
particles, a closed $N$-particle path
may induce a permutation
$P$ of the particles.
The trace includes a sum over all
permutations $P\in S_N$, which
reduces to a sum over all conjugation
classes ${\cal P}\subset S_N$.
Here $S_N$ is the symmetric group, and
the conjugation class ${\cal P}$
is characterized by a partition of $N$,
i.e.\ a sequence of non-negative integers
$\nu_1,\nu_2,\ldots$ such that
$\sum_L \nu_L L=N$.
Every permutation $P\in{\cal P}$
may be factored as a product of
commuting cycles with $\nu_L$ cycles
of length $L$, and the sign of $P$ is
$\sgn(P)=(-1)^{N-\nu}$, where
$\nu=\sum_L \nu_L$ is the number of
cycles. In Eq.\ (\ref{eq:ZNanyonI}),
${\cal C}({\cal P})$ is the set of
$N$-particle paths inducing an arbitrary,
but fixed permutation $P\in{\cal P}$.
$Q$ denotes the {\em winding number}
of the path ${\bmabit R}(\tau)$.
$Q$ is always an integer, it is even
when the permutation $P$ is even,
and is odd when $P$ is odd.

The path integral in
Eq.\ (\ref{eq:ZNanyonI})
with $\theta=0$ is
\beqn
\label{eq:pathintI}
\mathop{\int}_{{\cal C}({\cal P})}
{\cal D}({\bmabit R}(\tau))\,
e^{-{S\over\hbar}}=
\prod_L
\left(Z_1(L\beta)\right)^{\nu_L}.
\eeqn
This relation holds because
a factorization of the permutation
$P\in{\cal P}$ into disjoint cycles
implies a corresponding factorization
of the path integral, and
the path integral for one cycle of
length $L$ is $Z_1(L\beta)$.
In fact, the path integral for
$L$ bosons, when the permutation
is cyclic, equals the path integral
for one particle over $L$ times
as long a time interval.

Extracting from the path integral
a normalization factor given by
Eq.\ (\ref{eq:pathintI}), we write
\beqn
\label{eq:ZNanyonII}
Z_N(\beta,\theta)=
\sum_{\cal P}\left(
\prod_L {1\over\nu_L!}
\left({Z_1(L\beta)\over L}\right)
^{\nu_L}\right)F_{\cal P}(\beta,\theta),
\eeqn
where $F_{\cal P}$ is a probability
generating function,
\beqn
F_{\cal P}(\beta,\theta)=
\sum_{Q=-\infty}^{\infty}
P_{\cal P}(\beta,Q)\,e^{-i\theta Q}\;,
\eeqn
and $P_{\cal P}(\beta,Q)$ is the probability
of the winding number $Q$, given
the partition ${\cal P}$ and the weight
$e^{-S/\hbar}$ of each path. Since
$P_{\cal P}(\beta,-Q)=P_{\cal P}(\beta,Q)$,
for the reason that the probability
distribution of paths is time reversal
invariant, and since $Q$ is even/odd for
an even/odd permutation, the following
relations hold,
\beqn
F_{\cal P}(\beta,\theta)=
F_{\cal P}(\beta,-\theta)=
\sgn({\cal P})\,
F_{\cal P}(\beta,\pi-\theta).
\eeqn

In particular, the boson and fermion
values are $F_{\cal P}(\beta,0)=1$ and
$F_{\cal P}(\beta,\pi)=\sgn({\cal P})
=(-1)^{N-\nu}$.
Thus for $N$ non-interacting
bosons or fermions we have that
\beqn
Z_N(\beta)=
(\pm 1)^N
\sum_{\cal P}
\left(\prod_L
{1\over\nu_L!}\left(
\pm{Z_1(L\beta)\over L}
\right)^{\nu_L}\right),
\eeqn
with the upper sign for bosons and
the lower for fermions.
This relation is valid in any
dimension, as can be proved from
the grand canonical partition function,
where $E_n$ is the energy of the $n$-th
level in the one-particle system,
\beqn
\Xi(\beta,z)=\prod_n\left(
1\mp z e^{-\beta E_n}\right)^{\mp 1}
=\sum_{N=0}^{\infty}z^N Z_N(\beta).
\eeqn
The proof is by exponentiation of
the formula
\beqn
\ln(\Xi(\beta,z))=
\pm\sum_n \sum_{L=1}^{\infty}
{(\pm z)^L e^{-L\beta E_n}\over L}=
\sum_{L=1}^{\infty}
{(\pm z)^L\left(
\pm Z_1(L\beta)\right)\over L}\;.
\eeqn

\section{The Harmonic Oscillator}

In order to compute the virial coefficients,
it is convenient to confine the particles by
the harmonic oscillator potential
$U({\bmabit r})=m\omega^2|{\bmabit r}|^2/2$
and take the limit $\omega\to 0$.
The well-known one- and two-particle partition
functions depend on the dimensionless parameter
$\xi=\hbar\omega\beta$,
\beqn
Z_1(\beta)=
\left(2\sinh\!\left({\xi\over 2}
\right)\right)^{-2},
\eeqn
\beqn
Z_2(\beta,\theta)={1\over 2}
\left(Z_1(\beta)\right)^2
F_{11}(\beta,\theta)+
{1\over 2}
Z_1(2\beta)\,F_2(\beta,\theta)=
{\cosh\!\left(\left(
1-\alpha\right)\xi\right)\over
\left(2\sinh\!\left({\xi\over 2}
\right)\right)^2
2\sinh^2\!\xi}\;.
\eeqn
Here $\alpha=\alpha(\theta)$
is a ``sawtooth'' function,
$\alpha(\theta)=|\chi|/\pi$ when
$\theta=\chi+2n\pi$, $|\chi|\leq\pi$ and
$n$ is an integer.
We get for the even partition $1+1=2$ and
the odd partition $2=2$ that
\beqn
F_{11}(\beta,\theta)&\!\!\!=&\!\!\!
\sum_{\mbox{\scriptsize even }Q}
P_{11}(\beta,Q)\,e^{-i\theta Q}=
{\cosh\!\left(\left(
\alpha-{1\over 2}\right)\xi\right)
\over
\cosh\!\left({\xi\over 2}\right)}\;,
\nonumber\\
F_2(\beta,\theta)&\!\!\!=&\!\!\!
\sum_{\mbox{\scriptsize odd }Q}
P_2(\beta,Q)\,e^{-i\theta Q}=
-{\sinh\!\left(\left(
\alpha-{1\over 2}\right)\xi\right)
\over
\sinh\!\left({\xi\over 2}\right)}\;,
\eeqn
from which follows that
\beqn
\label{eq:Lorentzdist}
P_{11}(\beta,Q)=
{2\xi\tanh\!\left({\xi\over 2}\right)
\over\xi^2+(\pi Q)^2}\;,
\qquad
P_2(\beta,Q)=
{2\xi\coth\!\left({\xi\over 2}\right)
\over\xi^2+(\pi Q)^2}\;.
\eeqn

The three-anyon partition function is
\beqn
\label{eq:Zthreeanyon}
Z_3(\beta,\theta)={1\over 6}
\left(Z_1(\beta)\right)^3
F_{111}(\beta,\theta)+
{1\over 3}
Z_1(3\beta) F_3(\beta,\theta)+
{1\over 2}
Z_1(2\beta) Z_1(\beta)
F_{21}(\beta,\theta).
\eeqn
There are two even partitions, $1+1+1=3$
and $3=3$, and one odd, $2+1=3$.
The unknown parts of $Z_3$ are the probability
generating functions $F_{111}$,
$F_3$ and $F_{21}$, which we are
only able to calculate numerically,
except that the part of $Z_3$ which is
odd under the reflection $\theta\to\pi-\theta$
is known from the supersymmetry argument
of Sen \cite{SenI,SenII}. In our notation,
Sen's result is that
\beqn
\label{eq:F21expl}
F_{21}(\beta,\theta)=
F_2(3\beta,\theta)=
-{\sinh\left(\left(\alpha-{1\over 2}
\right)3\xi\right)\over
\sinh\left({3\xi\over 2}\right)}\;.
\eeqn

The distribution of winding numbers can be
found numerically by Monte Carlo
generation of paths according to an exact
multi-dimensional normal distribution.
For this we need the imaginary-time
one-particle propagator for
the two-dimensional oscillator,
\beqn
G({\bmabit s},{\bmabit r};\tau)
&\!\!\!=&\!\!\!
\langle{\bmabit s}|
e^{-(\tau/\hbar)\widehat{H}}|
{\bmabit r}\rangle\\
&\!\!\!=&\!\!\!
{m\omega\over
2\pi\hbar\sinh(\omega\tau)}
\exp\!\left(-{m\omega\over 4\hbar}\left(
\tanh\!\left({\omega\tau\over 2}\right)
|{\bmabit s}+{\bmabit r}|^2+
\coth\!\left({\omega\tau\over 2}\right)
|{\bmabit s}-{\bmabit r}|^2
\right)\right).
\nonumber
\eeqn
The common starting and ending point
${\bmabit r}={\bmabit r}(0)=
{\bmabit r}(L\hbar\beta)$
of a random closed path over the interval
$\tau=0$ to $\tau=L\hbar\beta$
has a probability density
proportional to
$G({\bmabit r},{\bmabit r};L\hbar\beta)$.
This defines a normal distribution
of mean zero and standard deviation
\beqn
\label{eq:sigma0}
\sigma_0=
\sqrt{{\hbar\over 2m\omega}
\coth\!\left({L\xi\over 2}
\right)}\;
\mathop{\longrightarrow}_{\omega\to 0}\;
{1\over\sqrt{mL\omega^2\beta}}\;.
\eeqn
Given three times $\tau_a<\tau<\tau_b$
and the two points
${\bmabit r}(\tau_a)={\bmabit r}_a$
and ${\bmabit r}(\tau_b)={\bmabit r}_b$
on the path, the position
${\bmabit r}(\tau)={\bmabit r}$
has a probability density proportional to
$G({\bmabit r}_b,{\bmabit r};\tau_b-\tau)\,
G({\bmabit r},{\bmabit r}_a;\tau-\tau_a)$.
This defines a normal distribution of mean
\beqn
\label{eq:rmeantau}
{\bmabit r}_{\tau}=
{\sinh(\omega(\tau_b-\tau))\,{\bmabit r}_a
+\sinh(\omega(\tau-\tau_a))\,{\bmabit r}_b\over
\sinh(\omega(\tau_b-\tau_a))}
\eeqn
and standard deviation
\beqn
\label{eq:sigmatau}
\sigma_{\tau}=\sqrt{\hbar\,
\sinh(\omega(\tau-\tau_a))\,
\sinh(\omega(\tau_b-\tau))\over m\omega\,
\sinh(\omega(\tau_b-\tau_a))}\;
\mathop{\longrightarrow}_{\omega\to 0}\;
\sqrt{\hbar(\tau-\tau_a)(\tau_b-\tau)
\over m(\tau_b-\tau_a)}\;.
\eeqn

We use these formulae to generate
random $N$-particle paths.
Given a value of $\beta$
and a partition $\sum_L \nu_L L=N$,
we treat one cycle at a time.
A cycle of length $L$ corresponds to
an $L$-particle path over an
imaginary-time interval
$\hbar\beta$, or equivalently a closed
one-particle path
over the interval $L\hbar\beta$.
We generate first the common starting
and ending point of this one-particle
path, and then a successively finer
subdivision of the $\tau$-interval.
We count the windings by counting
sign changes of the relative
coordinates.

\section{The Third Virial Coefficient}

The second and third virial coefficients
are, with
$\lambda=\hbar\sqrt{2\pi\beta/m}$
the thermal wave length,
\beqn
A_2(\theta)=\lim_{\omega\to 0}
\left(\lambda\over\xi\right)^2
\left(1-2{Z_2(\beta,\theta)\over(Z_1(\beta))^2}
\right)=
\lambda^2\left({1\over 4}-
{(1-\alpha)^2\over 2}\right),
\eeqn
\beqn
A_3(\theta)=\lim_{\omega\to 0}
\left(\lambda\over\xi\right)^4
\left(2-10{Z_2(\beta,\theta)
\over(Z_1(\beta))^2}+
16{(Z_2(\beta,\theta))^2\over(Z_1(\beta))^4}-
6{Z_3(\beta,\theta)\over(Z_1(\beta))^3}\right).
\eeqn

In the limit $\omega\to 0$,
Eq.\ (\ref{eq:sigma0}) implies that
the starting and ending point of
an $L$-particle loop is located inside
a region of area inversely proportional
to $\omega^2$. Furthermore,
Eq.\ (\ref{eq:sigmatau}) implies that
the area covered by the $L$-particle cycle
tends to a non-zero, finite limit
as $\omega\to 0$.
The probability that particles belonging to
two different cycles wind around each other,
is therefore proportional to $\omega^2$.
The probability that three cycles
overlap simultaneously, is
proportional to $\omega^4$.
These estimates imply for
the function $F_{111}$ that
the windings of the three pairs of
particles are uncorrelated up to
terms of order $\omega^4$. Hence,
\beqn
\label{eq:FaaaO}
F_{111}(\beta,\theta)=
\left(F_{11}(\beta,\theta)\right)^3
\left(1+F_{111}^{(4)}(\theta)\,\xi^4
+{\cal O}(\xi^5)\right).
\eeqn
Similarly, for the function $F_{21}$
only the two-cycle contributes to
the windings up to terms of order
$\omega^2$. Hence,
\beqn
\label{eq:FbaO}
F_{21}(\beta,\theta)=
F_2(\beta,\theta)
\left(1+F_{21}^{(2)}(\theta)\,\xi^2
+{\cal O}(\xi^3)\right).
\eeqn
The explicit formula in
Eq.\ (\ref{eq:F21expl}) is of this form,
with
\beqn
\label{eq:F212expl}
F_{21}^{(2)}(\theta)=-{1\over 3}
+{4\over 3}
\left(\alpha-{1\over 2}\right)^2.
\eeqn
We write also
\beqn
\label{eq:FcO}
F_3(\beta,\theta)=
F_3^{(0)}(\theta)+{\cal O}(\xi).
\eeqn

These expansions imply the non-trivial
result that $A_3(\theta)$ is finite
for all $\theta$. Explicitly it is
\beqn
A_3(\theta)&\!\!\!=&\!\!\!\lambda^4\left({1\over 64}
+{7\over 8}\left(\alpha-{1\over 2}\right)^2
+{1\over 4}\left(\alpha-{1\over 2}\right)^4
-F_{111}^{(4)}(\theta)
-{2\over 9}F_3^{(0)}(\theta)\right.
\nonumber\\&\!\!\!&\!\!\!\left.
+{1\over 2}\left(\alpha-{1\over 2}\right)
\left(3F_{21}^{(2)}(\theta)+1
-4\left(\alpha-{1\over 2}\right)^2\right)
\right).
\eeqn
Note that all the three functions
$F_{111}^{(4)}$, $F_3^{(0)}$ and
$F_{21}^{(2)}$ are even under the reflection
$\theta\to\pi-\theta$. Thus we see that $A_3$
is even if and only if
Eq.\ (\ref{eq:F212expl}) holds.

\section{Numerical Results and Discussion}

Due to statistical fluctuations,
the Monte Carlo estimates
$P_{\cal P}^{\mbox{\scriptsize MC}}(\beta,Q)$
for the probabilities
will usually violate the symmetry
$P(Q)=P(-Q)$, so that the estimated
probability generating function
$F_{\cal P}^{\mbox{\scriptsize MC}}
(\beta,\theta)$ becomes complex.
We may then use the imaginary part as a measure
of the statistical uncertainty in the real part.

For a given $\xi=\hbar\omega\beta$,
$F_{\cal P}^{\mbox{\scriptsize MC}}$ is exact at
$\theta=0$ and $\theta=\pi$. One way
to check it against some theoretical prediction
is therefore to check the derivative
with respect to $\theta$.
In Fig.\ \ref{figF21} is plotted
$\pi\partial F_{21}^{\mbox{\scriptsize MC}}/
\partial \theta$ versus $\theta/\pi$,
based on $10^5$ paths generated at $\xi=1$.
We plot only the interval
$0\leq\theta\leq\pi/2$, because of
the periodicity and symmetry relations.
The imaginary part is plotted in order
to indicate the statistical uncertainty.
The smooth curve is the derivative of
Eq.\ (\ref{eq:F21expl}), it is antisymmetric
about $\theta=0$ and obviously discontinuous.
In addition to statistical fluctuations,
there is a systematic deviation of
the Monte Carlo generated curve from
the exact curve. We interpret this as
an example of the Gibbs phenomenon, which
is unavoidable when a discontinuous function
is approximated by a finite Fourier series.
It is characteristic that the finite series
``overshoots'' close to the discontinuity.
Allowing for statistical and systematic
errors, we conclude from Fig.\ \ref{figF21}
that our Monte Carlo simulation agrees with
the theoretical result due to Sen,
Eq.\ (\ref{eq:F21expl}). Thus the comparison
serves to verify both.

Fig.\ \ref{figF3} shows
$\pi\partial F_3^{\mbox{\scriptsize MC}}/
\partial\theta$
as a function of $\theta/\pi$, together with
the straight line $9(\alpha-{1\over 2})$.
$2\times 10^5$ paths were generated at
$\omega=0$, i.e.\ in the exact
free-particle limit.
Based on this figure, in comparison with
Fig.\ \ref{figF21}, we conjecture that
the straight line is indeed the exact
derivative of $F_3$ in this
$\theta$-interval, which would mean that
\beqn
F_3^{(0)}(\theta)=-{1\over 8}+
{9\over 2}\left(\alpha-{1\over 2}\right)^2.
\eeqn
This formula is at least a very good
approximation, if not exact.

The most difficult quantity to extract
from Monte Carlo data is the fourth
order term $F_{111}^{(4)}(\theta)$.
Our result, based on $1.8\times 10^7$
paths generated at $\xi=0.25$,
is plotted in Fig.\ \ref{figF111}.
The imaginary part is also plotted and
shows that the statistical uncertainty
is relatively large. We have verified
the $\xi^4$ dependence and thus
the extrapolation to $\xi=0$ by
comparison with Monte Carlo simulations
at $\xi=0.5$ and $\xi=0.75$.
In the figure are also plotted
two curves corresponding to
the two curves in Fig.\ \ref{figa3}.

Fig.\ \ref{figa3} shows the numerically
calculated third virial coefficient
$A_3(\theta)$, compared to
Eq.~(\ref{eq:A3conjecture}).
One theoretical curve represents
the first two terms in
Eq.~(\ref{eq:A3conjecture}), with
no free parameter, the other curve is
a two-parameter least squares fit
and has $c_4=0.00658$, $c_6=-0.00583$.
We conclude from this figure and from
Fig.\ \ref{figF111} that the Fourier
components $c_4,c_6,\ldots$
are small, but we can not tell with
certainty whether or not
they vanish exactly.

One further comment is in place.
The asymptotic form $1/Q^2$ of
the Lorentz distributions of winding
numbers in Eq.\ (\ref{eq:Lorentzdist})
is typical of all winding number
distributions, thus the probability
of high winding numbers is not
negligible. In our Monte Carlo
simulation the high winding numbers
occur at extremely small distances,
in fact we simulate distances down
to $10^{-300}$, which is of course
entirely unrealistic. It would be
much more realistic to introduce
some hard core repulsion between
the anyons. This would cut off
the tails of the winding number
distributions and thus guarantee
that the $\theta$-dependence is
analytic. See e.g.\
\cite{AS:MKS:RKB:JL:secvirhd,DL:YF:secvirint}.

\section*{Acknowledgement}

We thank Per Arne Slotte for good
advice on random number generators.

\newpage

\begin{figure}
\vspace*{4in}

\makebox[4in][l]{
\hspace*{-0.4in}\special{fig1.ps}
}

\caption{Real and imaginary parts of
$\pi\partial F_{21}^{\mbox{\scriptsize MC}}/
\partial\theta$,
at $\xi=1$, versus $\theta/\pi$.
The smooth curve is the exact result
due to Sen.}
\label{figF21}
\end{figure}

\begin{figure}
\vspace*{4in}

\makebox[4in][l]{
\hspace*{-0.4in}\special{fig2.ps}
}

\caption{Real and imaginary parts of
$\pi\partial F_{3}^{\mbox{\scriptsize MC}}/
\partial\theta$,
at $\xi=0$, versus $\theta/\pi$.
The real part is consistent with
the straight line
$9(\alpha-{1\over 2})$.}
\label{figF3}
\end{figure}

\begin{figure}
\vspace*{4in}

\makebox[4in][l]{
\hspace*{-0.4in}\special{fig3.ps}
}

\caption{Real and imaginary parts of
$F_{111}^{(4)\mbox{\scriptsize MC}}$,
as computed at $\xi=0.25$, versus $\theta/\pi$.
The two curves correspond to the two
curves in Fig.\ \protect\ref{figa3}.}
\label{figF111}
\end{figure}

\begin{figure}
\vspace*{4in}

\makebox[4in][l]{
\hspace*{-0.4in}\special{fig4.ps}
}

\caption{The third virial coefficient
$A_3$ as a function of
$\theta/\pi$, compared to
Eq.\ (\protect\ref{eq:A3conjecture})
with either $c_4=c_6=\ldots=0$ or
$c_4=0.00658$, $c_6=-0.00583$.}
\label{figa3}
\end{figure}

\end{document}